\documentclass{nature}

\usepackage{lineno}
\usepackage{lmodern}
\usepackage{amsmath}
\usepackage{amssymb}
\usepackage{graphicx}
\usepackage{hyperref}
\usepackage{xspace}
\usepackage{caption}
\usepackage{mathptmx}
\usepackage{mathrsfs}
\usepackage{pdfpages}
\usepackage{mathtools} 
\usepackage{bm}

\usepackage{xr-hyper} 
\newcommand{\sms}{\ensuremath{\mathbf{S}}\xspace}

\newcommand{\HH}{\mathcal{H}}

\title{
Properties and dynamics of meron topological spin textures in the two-dimensional magnet CrCl$_3$ 
} 

\author{Mathias Augustin$^{1}$, Sarah Jenkins$^2$, Richard F. L. Evans$^2$, Kostya S. Novoselov$^{3,4}$ \& Elton J. G. Santos$^{5,6\dagger}$}

\makeatletter
\let\saved@includegraphics\includegraphics
\AtBeginDocument{\let\includegraphics\saved@includegraphics}
\renewenvironment*{figure}{\@float{figure}}{\end@float}
\makeatother

\begin{document}

\maketitle

\begin{affiliations}
\item School of Mathematics and Physics, Queen's University Belfast, BT7 1NN, UK
\item Department of Physics, The University of York, YO10 5DD, UK 
\item Department of Material Science \& Engineering, National University of Singapore, Block EA, 
9 Engineering Drive 1, 117575, Singapore 
\item Chongqing 2D Materials Institute, Liangjiang New Area, Chongqing 400714, China 
\item Institute for Condensed Matter Physics and Complex Systems, School of Physics and Astronomy, The University of Edinburgh, EH9 3FD, UK.  
\item Higgs Centre for Theoretical Physics,  The University of Edinburgh,  EH9 3FD,  UK \\ 
$^{\dagger}$Corresponding author: esantos@ed.ac.uk  
\end{affiliations}
 
\date{}



\section*{Abstract}

\begin{abstract}

{\rm Merons are nontrivial topological spin textures highly relevant for many phenomena in solid state physics.  Despite their importance,  direct observation of such vortex quasiparticles is scarce and has been limited to a few complex materials.  Here we show the emergence of merons and antimerons in recently discovered two-dimensional (2D) CrCl$_3$ at zero magnetic field.  We show their entire evolution from pair creation,  their diffusion over metastable domain walls,  and collision leading to large magnetic monodomains. 
Both quasiparticles are stabilized spontaneously during cooling at regions where 
in-plane magnetic frustration takes place.  Their dynamics is determined by the interplay between the strong in-plane dipolar interactions and the weak out-of-plane magnetic anisotropy stabilising a vortex core within a radius of 8$-$10 nm.  Our results push the boundary to what is currently known about non-trivial spin structures in 2D magnets and open exciting opportunities to control magnetic domains via topological quasiparticles. }

\end{abstract}

\noindent 

\section*{Introduction} 

The finding of magnetism in atomically thin vdW 
materials has attracted much recent interest\cite{firstCrI3,CrGeTe,Guguchiaeaat3672,Klein1218}. 
The strict 2D nature of the layers leads to unique physical properties 
ranging from stacking dependent interlayer magnetism\cite{Xiaodong19,Song:2019aa}, 
giant tunneling magnetoresistance\cite{Song1214,Wang:2018aa} 
and second harmonic generation\cite{Sun:2019aa}, up to electric field control of 
magnetic properties\cite{Mak18}. Of particular interest 
are topological spin excitations\cite{Fert:2013aa,Muhlbauer:2009aa}, e.g. merons, which 
are crucial to understand fundamental problems of chiral magnetic order\cite{Tokura18} and 
development of novel spintronic devices for information technologies\cite{Fert:2013aa}. 
Layered magnetic materials provide an ideal platform to investigate and harness this critical spin 
phenomenon as the genuine character of meron systems is intrinsically 2D and 
the integration of magnetic sheets in device heterostructures is a reality\cite{Novoselov19}. 

Here we demonstrate that monolayer CrCl$_3$ hosts merons and antimerons 
in its magnetic structure. Both quasiparticles are created naturally 
during zero-field cooling at low temperatures. 
We find that both spin textures 
are directly associated with the metastability of the magnetic 
domains on CrCl$_3$ induced by spin fluctuations.  
The merons and antimerons assume a random distribution throughout 
the surface creating a network of topological spin textures with no 
apparent lattice-order as observed in other materials\cite{Tokura18}. 
The different sites of the network can interact to each 
other leading to different type of collisions involving meron and 
antimerons occurring within a nanosecond time-scale. 
Our results indicate that the control of vortex and antivortex in CrCl$_3$
is also a driving force for the manipulation of magnetic domains which 
follows closely the annihilation process of the spin textures.


\section*{Results}

Our starting point is the following spin Hamiltonian: 
\begin{equation}
{\cal H} = -\sum_{ij} J_{ij} ({\bf S}_i \cdot {\bf S}_j) - \sum_{ij} \lambda_{ij} S_i^z S_j^z -\sum_i D_i \left({\bf S}_i \cdot {\bf e}_i\right)^{2} - \sum_{ij} K_{ij} \left({\bf S}_i \cdot {\bf S}_j \right)^2 - \sum_i \mu_i {\bf S}_i \cdot \left ({\bf B}_i + {\bf B}^{\textrm{dp}}_i \right )   
\label{BQ-vamp}
\end{equation}
where $\sms_{i}$ and $\sms_{j}$ are the localized magnetic 
moments on Cr atomic sites $i$ and $j$ which 
are coupled by pair-wise exchange interactions. 
$J_{ij}$ and $\lambda_{ij}=J^{xx,yy}_{l}-J^{zz}_{l}$ (where $l$ sets the 
nearest neighbours taken into account) 
are the isotropic and anisotropic bilinear (BL) exchanges, respectively, 
and $D_i$ is the single ion magnetic anisotropy. 
${\bf B}_i$ and 
${\bf B}^{\textrm{dp}}_i$ represent external and dipole 
magnetic field sources respectively. 
We considered in Eq.\ref{BQ-vamp} up to third nearest neighbours for $J_{ij}$ ($J_1-J_2-J_3$) and 
$\lambda_{ij}$ ($\lambda_1-\lambda_2-\lambda_3$) in the description of CrCl$_3$.
The fourth term in Eq.\ref{BQ-vamp} represents the biquadratic (BQ) 
exchange which involves the hopping of two or more electrons  
between two adjacent sites\cite{Slonczewski,Kartsev:2020aa}.  
Its strength is given by the constant $K_{ij}$, which is the 
simplest and most natural form of non-Heisenberg coupling. 
We recently found that several 2D magnets develop 
substantial BQ exchange in their magnetic properties 
which is critical to quantitatively describe 
important features such as Curie temperatures, thermal stability 
and magnon spectra\cite{Kartsev:2020aa}. 
The magnitude of $K_{ij}$ for CrCl$_3$ is 0.22 meV 
which is slightly smaller than the BL exchange for the first-nearest neighbors 
$J_1=1.28$ meV but too large to be ignored. 
In our implementation the BQ exchange is quite general and can be 
applied to any pair-wise exchange interaction of arbitrary range. 
All the parameters in Eq.\ref{BQ-vamp} are extracted  
from non-collinear ab initio simulations including spin-orbit coupling to determine 
the different components of $J_{ij}$, $\lambda_{ij}$ and $K_{ij}$ ($ij=xx, yy, zz$) 
as described in Ref.\cite{Kartsev:2020aa}. 
For atomistic spin dynamics we calculate the effective magnetic field $B^i$ arising from 
the BQ exchange interactions with components:  
\begin{eqnarray}
B_x^i &=& -\frac{1}{\mu_i}\frac{\partial \HH_{\mathrm{}}}{\partial S_x} = 2 K_{{ij}} S_x^j \left(S_x^i S_x^j + S_y^i S_y^j + S_z^i S_z^j\right) \nonumber \\ 
B_y^i &=& -\frac{1}{\mu_i}\frac{\partial \HH_{\mathrm{}}}{\partial S_y} = 2 K_{{ij}} S_y^j \left(S_x^i S_x^j + S_y^i S_y^j + S_z^i S_z^j\right) \nonumber \\ 
B_z^i &=& -\frac{1}{\mu_i}\frac{\partial \HH_{\mathrm{}}}{\partial S_z} = 2 K_{{ij}} S_z^j \left(S_x^i S_x^j + S_y^i S_y^j + S_z^i S_z^j\right)\mathrm{.}
\end{eqnarray}
This effective field is then included within the total field $\mathbf{B}_{\mathrm{eff}}$ 
describing the time evolution of each atomic spin using the stochastic 
Landau-Lifshitz-Gilbert (LLG) equation: 
\begin{equation}\label{LLG}
\frac{\partial \sms_{i}}{\partial t} = - \frac{\gamma}{(1+\lambda^{2})}[\sms_{i} \times \mathbf{B}_{\mathrm{eff}}^{i} + \lambda \sms_{i} \times (\sms_{i} \times \mathbf{B}_{\mathrm{eff}}^{i}) ]
\end{equation} 
where $\gamma$ is the gyromagnetic ratio. See Supplementary Section \ref{SI-sec:computational} for details. 
The dynamics of merons and antimerons is determined 
at different temperatures $T$ and magnetic fields $\mathbf{B}_{\mathrm{eff}}$ over a time scale of more than 40 ns since the cooling starts at $T>>T_{c}$, where $T_c$ 
is the Curie temperature, until it reaches 0 K within 2 ns (Supplementary Section \ref{SI-movies}). 
Figure \ref{fig1} shows that as this process occurs, the nucleation of several 
small areas with an out-of-plane spin polarization $S^z$
perpendicular to the easy-plane of CrCl$_3$ 
appeared naturally at $T<T_{c}$ (see Supplementary Figure \ref{SI-Tc} and Supplementary Movie S1). 
%
As the temperature reached 0 K, there is a clear formation of several 
notch structures which are created randomly all over the crystal without following any apparent pattern 
or preferential nucleation site (Fig. \ref{fig1}{c}). Indeed, the number of dark and 
bright spin textures formed during the cooling at zero field 
is even indicating some equilibrium on the different signs of $S^z$. 
A similar nucleation process also happened at finite magnetic 
field perpendicular to the surface 
within the range of 5$-$10 mT (Supplementary Figure \ref{SI-low-field}). 
Major modifications are observed at a magnitude 
of 50 mT as the field breaks the reflection 
symmetry of the layer, thereby leading to an asymmetry 
with respect of the polarization of the 
spin textures created (Fig.\ref{fig1}{d-f} and Supplementary Movie S2). 
That is, more dark spin complexes are nucleated which follow the direction of the 
out-of-plane external field. The darker background relative to the 
zero-field situation indicates that the spins gained an additional $S^z$ component. 
At fields close to 100 mT just one kind of spin polarization is observed 
throughout the spin textures which vanished completely for larger magnitudes (see Supplementary Figure \ref{SI-largeB} and Supplementary Movie S3). We performed a partial hysteresis 
calculation to estimate the critical magnetic field 
to switch all the spins in the system 
including those at the distinct textures 
(Supplementary Figure \ref{SI-partial-hysteresis}). We  
found a magnitude of 200 mT which is surprisingly large for 
the small single ion anisotropy but consistent with recent 
magnetometry measurements\cite{Mak19}. 


In order to identify the nature of the dark and bright spin structures, we analysed closely  
the different patterns formed on CrCl$_3$ at zero field and low temperatures 
(Figure \ref{fig2}). We could identify four main spin textures by looking at the spatial
distribution of $S^z$ as labelled B1$-$B4 (Fig. \ref{fig2}{\bf a}). 
Even though they show similar magnitudes of $S^z$ along $x$ or $y$ 
(Fig. \ref{fig2}{\bf b}) the same does not apply for 
components $S^x$ and $S^y$. That is, some shape anisotropy 
is observed at the in-plane distribution of the magnetization. 
We estimated an average radius of around 
14.5 nm and 15.7 nm along of $x-$ and $y-$displacement, respectively, 
despite of the spin notches considered (Supplementary Section \ref{SI-bubble-radius}). 
This indicates a more elliptical pattern where the main distinction 
between the spin textures B1$-$B4 is the orientation of the spins 
from the core (out-of-plane)
to in-plane away from the centre forming a vortex structure. 
Such vortex structures are typical of non-trivial topological spin textures such 
skyrmions and anti-skyrmions\cite{Nagaosa:2013aa}. 
To unveil their true nature we can calculate the topological 
number $N$ given by\cite{Zang16,PhysRevB.91.224407,PhysRevB.83.100408,DasSarma08}: 

\begin{equation}\label{topological-N}
N = \frac{1}{4\pi}\int  \bm{\hat{s}} \cdot \left(\frac{\partial\bm{\hat {s}}}{\partial x} \times\frac{\partial \bm{\hat{s}}}{\partial y}  \right) d x ~d y 
\end{equation}
where $\hat{\bm s}=\frac{\bm s}{|\bm s|}$ is the three-component 
spin field. Eq.\ref{topological-N} indicates a product between the 
vorticity of the spin textures, which is determined by the  
direction of the in-plane components of the magnetic 
moment, e.g. $ \left(\frac{\partial\bm{\hat {s}}}{\partial x} \times\frac{\partial \bm{\hat{s}}}{\partial y}  \right)$, 
and the out-of-plane component of $\bm{\hat {s}}$. The latter is observed 
at the core of the vortex while the former near the perimeter  
aligning with the easy-plane of the material producing 
magnetic helicity and polarity. 
The obtained spin textures in the spin dynamics follow 
this description forming two types of distributions of 
the magnetization in terms of vortex (Fig.\ref{fig2}{\bf c, f}) and 
antivortex (Fig.\ref{fig2}{\bf d, e}). 
To characterize which kind of spin quasiparticle 
is present in CrCl$_3$ (e.g. skyrmions or merons), we evaluated 
the integral in Eq.\ref{topological-N} numerically for each topological spin textures 
to quantify $N$ (see Supplementary Section \ref{SI-sec:Topological-N}). 
This integral is performed under an area sufficiently large to include the totality of the 
spin-textures which was converged to $20~ nm \times 20 ~nm$. This ensures that no 
spin contributions are left outside of Eq. \ref{topological-N}.  All three components of 
$\bm{\hat {s}}$ are considered in the integration. 
We found that $N=\pm 1/2$ which is indicative of merons ($N=-1/2$) and 
antimerons ($N=+1/2$). Such fractional values of 
$N$ implicate that the total charge $Q=eN$ (where $e$ is the 
electron charge) of the merons and antimerons 
is also fractional $Q=\pm e/2$\cite{DasSarma08} with 
$N=\pm 1/2$ being a 
topological invariant\cite{DasSarma08} since a 
meron topologically has half the spin winding of a skyrmion. 
Indeed, the spin polarization of the core can also be used 
to identify core-up or core-down vortex and antivortex which  
differs from skyrmions and antiskyrmions\cite{Nagaosa:2013aa}. 
This gives the possibility to identify additional degrees of freedom 
for magnetic helicity and polarity\cite{Nagaosa:2013aa}. 

One of the main implications for spin textures to have a well defined 
topological number is that they should 
keep their magnetic ordering regardless external perturbations\cite{Nagaosa:2013aa} 
or even form a lattice of stable vortex-antivortex spin textures\cite{PhysRevB.91.224407,Tokura18}. 
Nevertheless, by observing the dynamics of the vortex and antivortex  
in CrCl$_3$ at longer times (Supplementary Movies S1, S2), we notice that 
both spin quasiparticles interact and disappeared through collisions 
which happened roughly 15$-$20 ns after they are created below 5 K. 
This lifetime is at least two orders of magnitude larger than that previously observed 
for merons and antimerons in thin ferromagnetic iron layer\cite{PhysRevLett.118.097203}, 
kagome magnet\cite{Pereiro:2014aa} and ferromagnetic permalloy disks\cite{Yimei18}
which suggested more stability for measurements. 
Figure \ref{fig3} shows that the collision process is independent of the core polarization of the 
meron or antimeron considered (Fig. \ref{fig2}{\bf d-g}) 
but it involves at least one vortex and one antivortex during the process. 
That is, collisions between pairs of the same type, 
such as vortex-vortex or antivortex-antivortex are not observed at 
any temperature and magnetic field which is in agreement 
with magnetic vortice theories\cite{Hubert-book}. 
Intriguingly, the vortex and antivortex with different core 
polarizations (Fig. \ref{fig3}{\bf a-c}) 
approach each other in spiral orbits 
rotating the in-plane magnetization to out-of-plane. This stabilizes  
new cores for both spin textures following the original 
core polarization at large separations. At close distances 
(roughly 2 lattice constants) both vortex and antivortex reduced gradually 
dissolving the complicated spin arrangement 
into the easy-plane of CrCl$_3$ with an emission of a spin-wave\cite{Koog05,PhysRevLett.97.177202} (Supplementary Movie S4). In the case of merons and antimerons 
with similar core polarization (Fig. \ref{fig3}{\bf d-f} and Supplementary Movie S5), there is an 
enlargement of the area with out-of-plane magnetization 
as both topological quasiparticles 
approach each other. This induces the formation of a 
composed spin texture with a larger extension over the 
surface (Fig. \ref{fig3}{\bf e}) which 
incidentally dissolves to an in-plane orientation at 
later times (Fig. \ref{fig3}{\bf f}). 
Both dynamics involving merons and antimerons of 
parallel or antiparallel core polarization are intrinsically coupled 
to the domain structure present in CrCl$_3$ being formed shortly 
($\sim0.1-0.2$~ns) after the formation of the magnetic domains. 
Figure \ref{fig3}{\bf g-j} and Supplementary Movie S6 
show a broad perspective of the magnetic domains as the 
collisions happen involving different spin textures (small dots). 
We noticed that both merons and antimerons are localized at the boundary 
between magnetic domains with different spin 
polarization (e.g. $S^z=+1$ and $S^z=-1$) where their displacement is 
along of the domain edges as $S^y$ is zero. 
We have not observed any factor or physical ingredient that 
would help to determine the distance between 
the merons and antimerons as they appear spontaneously in the zero-field cooling. 
The distance between the spin-textures in this sense follows the spin dynamics 
of the system being stochastic as our calculations indicated.
Similar configurations are also 
observed at another in-plane component of the magnetization, e.g. $S^x=0$, where the 
topological spin textures are localized (see Supplementary Figure \ref{SI-Sx-frustration}). 
We can understand this type of magnetic frustration in terms of the competing exchange 
interactions between the different nearest neighbours (1st, 2nd, 3rd) in CrCl$_3$\cite{Kartsev:2020aa}. 
In terms of the isotropic exchange ($J_{1st}$, $J_{2nd}$, $J_{3rd}$), 
we observe that the frustration takes places due to the third nearest-neighbour 
which has an anti-ferromagnetic exchange ($J_{3rd} =-0.025$ meV) relative 
to the first- ($J_{1st}=1.28$ meV) and the second-nearest neighbours 
($J_{2nd}= 0.072$ meV). 
In terms of the anisotropic exchange ($\lambda_{1st}$, $\lambda_{2nd}$, $\lambda_{3rd}$), 
the competition between second- and 
third-nearest neighbours ($\lambda_{2nd} =$ -0.0097 meV , $\lambda_{3rd}= -0.0051$ meV) 
with the first-nearest neighbours ($\lambda_{1st} = 0.020$ meV) 
induced that the in-plane spins ($S^x$, $S^y$) become negligible. 
In addition, a full in-plane spin polarization along $S^x$ and $S^y$ 
would lead to a singularity of the exchange energy which is avoided as $S^z$ 
becomes non-negligible\cite{PhysRevLett.97.177202}. 
This indicates that the intersections where both in-plane components of the 
magnetization converge to zero are an efficient environment 
to localize non-trivial topological spin textures. 
Such localization of magnetic vortices are typically seen in singly connected 
samples with the magnetic flux occurring at the 
junctions of magnetic domains. For instance, in permanent 
magnets\cite{Choe420,Templeton97,Goodenough58} and 
in soft-magnetic nanodisks\cite{Weiss00} which strong cross-tie domain 
wall structures and geometry play a critical role in the generation of the vortices. 
Nevertheless, we find that this is not the case 
for monolayer CrCl$_3$. There are no structural constraints in the 
stabilization of the magnetic domains and 
the low magnetic anisotropy would orientate 
the spins more freely without a preferential 
orientation within the easy-plane\cite{2LCrCl3}. 
Therefore, such spontaneous formation of merons and antimerons   
is an astonishing, previously unreported phenomena in the 
magnetism of any 2D vdW magnet.

An intriguing question that raised by the presence of 
these topological spin textures in CrCl$_3$ is what their physical origin. 
It is known that strongly inhomogeneous magnetic textures can be created 
due to the competition between local interactions, e.g. exchange 
and magnetic anisotropy, and long-range interactions 
mediated by demagnetizing fields and 
magnetic dipoles\cite{Fukamichi02,Doring68,Nagaosa:2013aa,dipole1,dipole2}. 
In the case of CrCl$_3$, the interplay between dipolar interactions and 
magnetic anisotropy is one of the main ingredients in the creation of 
merons and antimerons. Figure \ref{fig4}{\bf a-b} show that as the cooling process takes place, 
the different spin textures intrinsically carry a large component of the dipole moment 
perpendicular to the surface ($\pm 300$~mT). This is mainly localized at the core of 
the quasiparticle and assists in stabilizing a strong component of $S^z$. 
Indeed, there is almost no difference between the projection 
of the magnetization perpendicular to the surface and 
that for the dipole-field along $z$ (Fig. \ref{fig4}{\bf a-b}). This indicates a 
close relationship between dipole-dipole interactions and magnetism in CrCl$_3$. 
The distinct spin polarization of the core of the merons and antimerons  
does not give any significant variation on the magnitude of the 
dipole-field, which is larger than those within the in-plane 
components (Fig. \ref{fig4}{\bf c-d}). Surprisingly, both 
components of the dipole-field ($x$ and $y$) reach 
smaller magnitudes ($\pm 200$~mT) than those along $z$ throughout 
the surface, and are strictly zero at the position of the spin textures.

In this context, the in-plane dipole fields favour an in-plane magnetization 
whereas the single ion anisotropy orients the spins perpendicular 
to the surface. In CrCl$_3$ the magnetic anisotropy is small which 
allows most of the spins to follow the dipolar directions except those 
at the core of the merons and antimerons. In such particular locations, 
the stronger $z$ component of the dipole-field pushes the spins 
out-of-plane enhancing the magnetic anisotropy. In any other 
part of the surface without the topological spin textures, 
the magnetization rotates parallel to the 
surface as an effect of the dipolar interactions\cite{2LCrCl3}. It is worth 
mentioning that no transition between a previously stabilized magnetic configuration, 
such as stripes, into the non-trivial magnetic textures is noticed in the 
spin dynamics with or without an applied field as it has been suggested 
as a potential origin of bubbles or skyrmions\cite{Doniach82,Giess73,Takao83}. 
Moreover, we do not take into account asymmetric exchange  
(Dzyaloshinskii-Moriya interactions) into our simulations which was initially 
checked to have no effect on the dynamics of the merons and 
antimerons (Supplementary Figure \ref{SI-DMI_2nd}). 
This excludes additional mechanisms 
based on relativistic effects\cite{Nagaosa:2013aa}. 
We also considered simulations without 
the inclusion of dipolar fields (Supplementary Figure \ref{SI-no_dipole}) for CrCl$_3$
with a two-fold implication. First, 
there is no appearance of non-trivial topological spin textures as 
the magnetization is consistently out-of-plane over the entire crystal. 
Second, there is the stabilization of an easy-axis perpendicular to the surface 
following the single ion anisotropy. That is, there is a suppression 
of the easy-plane experimentally observed for CrCl$_3$\cite{2LCrCl3}. 
Even though other models\cite{dipole1,dipole2} utilized for magnetic 
thin films assumed that the dipole-dipole directions can be effectively 
replaced by an easy-plane (XY), our calculations indicate that the inclusion of  
dipolar interactions plays a key role in the description 
of the magnetic properties of CrCl$_3$ (Supplementary Figure \ref{SI-textures}). 
The spins textures formed at such artificial easy-plane\cite{dipole1,dipole2} 
looked more chaotic than those computed without 
such restriction being more complex to assign any clear feature or 
to determine a topological number $N$.

We also noticed that the inclusion of biquadratic exchange and next-nearest 
neighbours is critical for the stabilization of non-trivial topological spin textures in 
2D magnets (see Supplementary Figures \ref{SI-No_next_neighbours}, \ref{SI-No_BQ}). 
For the former, the higher-order exchange between the spins gives further 
stability to Eq.\ref{BQ-vamp} since the sign of $K_{ij}=0.22$~meV is 
positive\cite{Kartsev:2020aa}. For the latter, the non-inclusion of second- and third-nearest neighbours
in Eq. \ref{BQ-vamp} resulted in the absence of vortex or antivortex spin-textures even if 
dipolar-interactions are considered. This can be reasoning in terms of 
the large contributions of the in-plane anisotropic exchange 
from the second- and third-nearest neighbours ($\lambda_{2nd}+\lambda_{3rd}=-10.25~ \mu$eV) 
relative to the first-nearest neighbours ($\lambda_{1st}=20.07~ \mu$eV). 
This indicates that although the dipolar interactions can assist in 
the creation of an easy-plane over the surface, they are not 
sufficient to polarize an in-plane magnetization ($S^{x}$, $S^{y}$) 
which is greatly affected by the next-nearest neighbour 
exchange interactions. 
Furthermore, we have extended our simulations for other materials in the Cr-trihalide family 
CrX$_3$, X$=$F, Br, I (see Supplementary Section \ref{SI-sec:Cr-halide} 
and Supplementary Movies S7-S9). This will allow us to  
have a broader perspective whether other layers 
in the same family, and in a similar chemical environment (e.g. halogens) and 
symmetry (e.g. honeycomb), would be susceptible for the creation of 
merons and antimerons. We found no evidence of non-trivial spin textures as those in 
CrCl$_3$. There is a continuous evolution of the magnetic domains 
from high temperatures till 0 K with no formation of complex spin configurations. 
We can understand these results in terms of the strong 
out-of-plane single anisotropy\cite{Kartsev:2020aa} 
present in CrX$_3$ (X$=$F, Br, I) with the dipolar interactions 
also pointing perpendicular to the surface. This indicates that the 
presence of an easy-plane is an important requisite for the stabilization of non-trivial 
spin quasiparticles. 

The complex dynamics of the merons and antimerons in CrCl$_3$ can be directly related to the 
magnetic stability of the layer\cite{Hubert-book}. 
Strong thermally-driven spin fluctuations 
prevent a clear observation of the distinct spin textures which are well 
pronounced for $T\leq5$ K (Figure \ref{fig1}). 
Such spin fluctuations are due in part 
to the low single ion anisotropy ($D\sim 0.01$ meV) of CrCl$_3$ 
which would require low-energy excitations to change the orientation  
of the surface spins. 
Other contribution factor is the metastability of the magnetic domains 
in CrCl$_3$ that continuously evolve as a function of time. 
For magnetic compounds 
with strong easy-plane anisotropy, demagnetization or long-range dipolar fields are 
known to affect the ground-state of 
topological spin-textures\cite{Vousden:2016aa,Beg:2015aa} which 
an equilibrium state is normally achieved beyond the field cooling process ending at 0 K. 
Although any thermal contribution to the magnetic domains will be zero at this limit,  
the spins would still evolve to stabilize the ground-state via the 
minimization of other contributions of the total 
energy, e.g. exchange, anisotropy. 
This process can be 
observed in Figure \ref{fig5}{\bf a,} for the time-evolution 
of one of the spin dynamics of monolayer CrCl$_3$ once the system 
had achieved 0 K within 2.0 ns at zero field. 
There is a continuous modification of the domain-wall profiles through all components of the 
magnetization ($S^{x}$, $S^y$, $S^z$) over time. The variations on $S^{x}$ and $S^{y}$ 
across the magnetic domains (Fig. \ref{fig5}{\bf b-c}) tend to 
be broader with less peaky changes as those observed 
along $S^z$ due to the presence of merons and antimerons (Fig. \ref{fig5}{\bf d}). 
For them, several sharp changes 
appeared and vanished on a time scale of few tenths of nanoseconds indicating 
the stochastic nature of the spin-fluctuations in the system. Indeed, we observed 
such random fluctuations of the magnetization even beyond 4 ns which suggests 
that the system may not be in a local minimum but rather at a
flat energy landscape. As a matter of fact, an increment of $D$ will not 
cease the fluctuations as our calculations showed that they may be intrinsic to 
CrCl$_3$ (Supplementary Figure \ref{SI-36mueV}).

\section*{Discussion}

The discovery of non-trivial topological spin-textures (merons and antimerons) 
in a non-chiral 2D magnetic material (CrCl$_3$) opens the possibility 
for other layered materials 
display such behaviour.  
Some guidelines for looking into 
materials that may develop 
such quasiparticles would be $i)$ a weak out-of-plane single-ion 
anisotropy ($D_i$),  $ii)$ high in-plane 
dipolar interactions,  and $iii)$ competition between next-nearest 
neighbours.  Our simulations indicate that 
the general nature of the formation of merons and anti-merons in a 
2D magnet is due to the combination of these three factors as cooling occurs. 
The delicate balance between anisotropy,  dipole-dipole interactions and exchange 
competition has a major effect in the 
stabilization of the core vortex,  the perimeter and the spin helicity as well as 
the polarity of the non-collinear spin textures. 
Our simulation results also indicate the observation of meron and antimeron 
spin textures at no applied magnetic field,  low-temperatures and 
without edge effects\cite{Nagaosa:2013aa,Nagaosa13}.  
Thus,  the problem now turns in the search of other layered 
compounds where such guidelines could be 
fulfilled.  Efforts on the discovery of novel magnetic sheets that 
hold topological non-trivial quasiparticles will pass 
through the accurate calculation of magnetic parameters, 
and subsequent atomistic simulation of large scale 
properties (i.e.  magnetic domains,  domain walls).   
Furthermore,  interactions with a substrate are also important to be considered. 
Even though we have not considered them explicitly in our calculations, 
different support could either enhance or deteriorate the 
magnetic ordering of thin layered materials.  
Progress in the isolation of CrCl$_3$ from 
spurious support interactions have recently been 
reported\cite{Amilcar20}.  Such experiments where large scale 
monolayer CrCl$_3$ are grown via molecular beam epitaxy 
on graphene/6H-SiC(0001) substrates provided a new avenue for the 
validation of the predictions included in this study.  
In addition,  merons and anti-merons are likely to be created under ultrafast 
laser excitations\cite{Yimei18} and current pulses\cite{Buttner:2017aa} which open 
the door for validating our predictions.  
With the prompt integration of magnetic layered materials in device platforms and the 
emergence of more compounds with similar characteristics,  it is a matter of 
time till experimental realization and subsequent control will be achieved 
for such non-trivial spin topology.  In this sense,  the emergent electrodynamics initially 
established for skyrmions\cite{Schulz:2012aa} can be explored further at a more 
fundamental level using merons in a truly 2D magnet.

\section*{Methods}
\label{sec:org3881bef}

All methods are included in Supplementary Information which includes Supplementary Sections S1$-$S5, Supplementary Movies S1$-$S9 and Supplementary Figures S1$-$S14.


\subsubsection*{Data Availability} 

The data that support the findings of this study 
are available within the paper and its Supplementary Information.  

\subsubsection*{Competing interests}
The Authors declare no conflict of interests.

\subsubsection*{Acknowledgments}
EJGS thanks Dina-Abdul Wahab for assistance in the preparation of Figures 5, S8-S10. 
RFLE gratefully acknowledges the financial support of the Engineering and Physical Sciences 
Research Council (Grant No. EPSRC EP/P022006/1) and the use of the VIKING Cluster, 
which is a high performance compute facility provided by the University of York. 
This work was enabled by code enhancements to the VAMPIRE software 
implemented under the embedded CSE programme (ecse0709) and (ecse1307) 
of the ARCHER UK National Supercomputing Service. 
EJGS also acknowledges computational resources through the 
UK Materials and Molecular Modeling Hub for access to THOMAS supercluster, 
which is partially funded by EPSRC (EP/P020194/1); CIRRUS Tier-2 HPC 
Service (ec131 Cirrus Project) at EPCC (http://www.cirrus.ac.uk) funded 
by the University of Edinburgh and EPSRC (EP/P020267/1); 
ARCHER UK National Supercomputing Service (http://www.archer.ac.uk) via 
Project d429. EJGS acknowledges the 
EPSRC Early Career Fellowship (EP/T021578/1) and 
the University of Edinburgh for funding support.

\subsubsection*{Author Contributions} 
EJGS conceived the idea and supervised the project. 
MA performed ab initio and Monte Carlo simulations 
under the supervision of EJGS. SJ implemented 
the dipole approximations. MA and EJGS elaborated 
the analysis with inputs from RFLE and KSN.  
EJGS wrote the paper with inputs from all authors.  
All authors contributed to this work, 
read the manuscript, discussed the results, and agreed 
to the contents of the manuscript.

\section*{References}

\begin{thebibliography}{10}
\expandafter\ifx\csname url\endcsname\relax
  \def\url#1{\texttt{#1}}\fi
\expandafter\ifx\csname urlprefix\endcsname\relax\def\urlprefix{URL }\fi
\providecommand{\bibinfo}[2]{#2}
\providecommand{\eprint}[2][]{\url{#2}}

\bibitem{firstCrI3}
\bibinfo{author}{Huang, B.} \emph{et~al.}
\newblock \bibinfo{title}{Layer-dependent ferromagnetism in a van der waals
  crystal down to the monolayer limit}.
\newblock \emph{\bibinfo{journal}{Nature}} \textbf{\bibinfo{volume}{546}},
  \bibinfo{pages}{270 EP --} (\bibinfo{year}{2017}).
\newblock \urlprefix\url{https://doi.org/10.1038/nature22391}.

\bibitem{CrGeTe}
\bibinfo{author}{Gong, C.} \emph{et~al.}
\newblock \bibinfo{title}{Discovery of intrinsic ferromagnetism in
  two-dimensional van der waals crystals}.
\newblock \emph{\bibinfo{journal}{Nature}} \textbf{\bibinfo{volume}{546}},
  \bibinfo{pages}{265--269} (\bibinfo{year}{2017}).
\newblock \urlprefix\url{http://dx.doi.org/10.1038/nature22060}.
\newblock \bibinfo{note}{Letter}.

\bibitem{Guguchiaeaat3672}
\bibinfo{author}{Guguchia, Z.} \emph{et~al.}
\newblock \bibinfo{title}{Magnetism in semiconducting molybdenum
  dichalcogenides}.
\newblock \emph{\bibinfo{journal}{Science Advances}}
  \textbf{\bibinfo{volume}{4}} (\bibinfo{year}{2018}).
\newblock
  \urlprefix\url{https://advances.sciencemag.org/content/4/12/eaat3672}.
\newblock
  \eprint{https://advances.sciencemag.org/content/4/12/eaat3672.full.pdf}.

\bibitem{Klein1218}
\bibinfo{author}{Klein, D.~R.} \emph{et~al.}
\newblock \bibinfo{title}{Probing magnetism in 2d van der waals crystalline
  insulators via electron tunneling}.
\newblock \emph{\bibinfo{journal}{Science}} \textbf{\bibinfo{volume}{360}},
  \bibinfo{pages}{1218--1222} (\bibinfo{year}{2018}).
\newblock \urlprefix\url{http://science.sciencemag.org/content/360/6394/1218}.
\newblock
  \eprint{http://science.sciencemag.org/content/360/6394/1218.full.pdf}.

\bibitem{Xiaodong19}
\bibinfo{author}{Chen, W.} \emph{et~al.}
\newblock \bibinfo{title}{Direct observation of van der waals
  stacking{\textendash}dependent interlayer magnetism}.
\newblock \emph{\bibinfo{journal}{Science}} \textbf{\bibinfo{volume}{366}},
  \bibinfo{pages}{983--987} (\bibinfo{year}{2019}).
\newblock \urlprefix\url{https://science.sciencemag.org/content/366/6468/983}.
\newblock
  \eprint{https://science.sciencemag.org/content/366/6468/983.full.pdf}.

\bibitem{Song:2019aa}
\bibinfo{author}{Song, T.} \emph{et~al.}
\newblock \bibinfo{title}{Switching 2d magnetic states via pressure tuning of
  layer stacking}.
\newblock \emph{\bibinfo{journal}{Nature Materials}}
  \textbf{\bibinfo{volume}{18}}, \bibinfo{pages}{1298--1302}
  (\bibinfo{year}{2019}).
\newblock \urlprefix\url{https://doi.org/10.1038/s41563-019-0505-2}.

\bibitem{Song1214}
\bibinfo{author}{Song, T.} \emph{et~al.}
\newblock \bibinfo{title}{Giant tunneling magnetoresistance in spin-filter van
  der waals heterostructures}.
\newblock \emph{\bibinfo{journal}{Science}} \textbf{\bibinfo{volume}{360}},
  \bibinfo{pages}{1214--1218} (\bibinfo{year}{2018}).
\newblock \urlprefix\url{http://science.sciencemag.org/content/360/6394/1214}.
\newblock
  \eprint{http://science.sciencemag.org/content/360/6394/1214.full.pdf}.

\bibitem{Wang:2018aa}
\bibinfo{author}{Wang, Z.} \emph{et~al.}
\newblock \bibinfo{title}{Very large tunneling magnetoresistance in layered
  magnetic semiconductor cri3}.
\newblock \emph{\bibinfo{journal}{Nature Communications}}
  \textbf{\bibinfo{volume}{9}}, \bibinfo{pages}{2516} (\bibinfo{year}{2018}).
\newblock \urlprefix\url{https://doi.org/10.1038/s41467-018-04953-8}.

\bibitem{Sun:2019aa}
\bibinfo{author}{Sun, Z.} \emph{et~al.}
\newblock \bibinfo{title}{Giant nonreciprocal second-harmonic generation from
  antiferromagnetic bilayer cri3}.
\newblock \emph{\bibinfo{journal}{Nature}} \textbf{\bibinfo{volume}{572}},
  \bibinfo{pages}{497--501} (\bibinfo{year}{2019}).
\newblock \urlprefix\url{https://doi.org/10.1038/s41586-019-1445-3}.

\bibitem{Mak18}
\bibinfo{author}{Jiang, S.}, \bibinfo{author}{Shan, J.} \&
  \bibinfo{author}{Mak, K.~F.}
\newblock \bibinfo{title}{Electric-field switching of two-dimensional van der
  waals magnets}.
\newblock \emph{\bibinfo{journal}{Nature materials}} \bibinfo{pages}{1}
  (\bibinfo{year}{2018}).

\bibitem{Fert:2013aa}
\bibinfo{author}{Fert, A.}, \bibinfo{author}{Cros, V.} \&
  \bibinfo{author}{Sampaio, J.}
\newblock \bibinfo{title}{Skyrmions on the track}.
\newblock \emph{\bibinfo{journal}{Nature Nanotechnology}}
  \textbf{\bibinfo{volume}{8}}, \bibinfo{pages}{152--156}
  (\bibinfo{year}{2013}).
\newblock \urlprefix\url{https://doi.org/10.1038/nnano.2013.29}.

\bibitem{Muhlbauer:2009aa}
\bibinfo{author}{M{\"u}hlbauer, S.} \emph{et~al.}
\newblock \bibinfo{title}{Skyrmion lattice in a chiral magnet}.
\newblock \emph{\bibinfo{journal}{Science}} \textbf{\bibinfo{volume}{323}},
  \bibinfo{pages}{915} (\bibinfo{year}{2009}).
\newblock
  \urlprefix\url{http://science.sciencemag.org/content/323/5916/915.abstract}.

\bibitem{Tokura18}
\bibinfo{author}{Yu, X.~Z.} \emph{et~al.}
\newblock \bibinfo{title}{Transformation between meron and skyrmion topological
  spin textures in a chiral magnet}.
\newblock \emph{\bibinfo{journal}{Nature}} \textbf{\bibinfo{volume}{564}},
  \bibinfo{pages}{95--98} (\bibinfo{year}{2018}).
\newblock \urlprefix\url{https://doi.org/10.1038/s41586-018-0745-3}.

\bibitem{Novoselov19}
\bibinfo{author}{Gibertini, M.}, \bibinfo{author}{Koperski, M.},
  \bibinfo{author}{Morpurgo, A.~F.} \& \bibinfo{author}{Novoselov, K.~S.}
\newblock \bibinfo{title}{Magnetic 2d materials and heterostructures}.
\newblock \emph{\bibinfo{journal}{Nature Nanotechnology}}
  \textbf{\bibinfo{volume}{14}}, \bibinfo{pages}{408--419}
  (\bibinfo{year}{2019}).
\newblock \urlprefix\url{https://doi.org/10.1038/s41565-019-0438-6}.

\bibitem{Slonczewski}
\bibinfo{author}{Slonczewski, J.~C.}
\newblock \bibinfo{title}{Fluctuation mechanism for biquadratic exchange
  coupling in magnetic multilayers}.
\newblock \emph{\bibinfo{journal}{Phys. Rev. Lett.}}
  \textbf{\bibinfo{volume}{67}}, \bibinfo{pages}{3172--3175}
  (\bibinfo{year}{1991}).
\newblock \urlprefix\url{https://link.aps.org/doi/10.1103/PhysRevLett.67.3172}.

\bibitem{Kartsev:2020aa}
\bibinfo{author}{Kartsev, A.}, \bibinfo{author}{Augustin, M.},
  \bibinfo{author}{Evans, R. F.~L.}, \bibinfo{author}{Novoselov, K.~S.} \&
  \bibinfo{author}{Santos, E. J.~G.}
\newblock \bibinfo{title}{Biquadratic exchange interactions in two-dimensional
  magnets}.
\newblock \emph{\bibinfo{journal}{npj Computational Materials}}
  \textbf{\bibinfo{volume}{6}}, \bibinfo{pages}{150} (\bibinfo{year}{2020}).
\newblock \urlprefix\url{https://doi.org/10.1038/s41524-020-00416-1}.

\bibitem{Mak19}
\bibinfo{author}{Kim, H.~H.} \emph{et~al.}
\newblock \bibinfo{title}{Evolution of interlayer and intralayer magnetism in
  three atomically thin chromium trihalides}.
\newblock \emph{\bibinfo{journal}{Proceedings of the National Academy of
  Sciences}} \textbf{\bibinfo{volume}{116}}, \bibinfo{pages}{11131--11136}
  (\bibinfo{year}{2019}).
\newblock \urlprefix\url{https://www.pnas.org/content/116/23/11131}.
\newblock \eprint{https://www.pnas.org/content/116/23/11131.full.pdf}.

\bibitem{Nagaosa:2013aa}
\bibinfo{author}{Nagaosa, N.} \& \bibinfo{author}{Tokura, Y.}
\newblock \bibinfo{title}{Topological properties and dynamics of magnetic
  skyrmions}.
\newblock \emph{\bibinfo{journal}{Nature Nanotechnology}}
  \textbf{\bibinfo{volume}{8}}, \bibinfo{pages}{899--911}
  (\bibinfo{year}{2013}).
\newblock \urlprefix\url{https://doi.org/10.1038/nnano.2013.243}.

\bibitem{Zang16}
\bibinfo{author}{Yin, G.} \emph{et~al.}
\newblock \bibinfo{title}{Topological charge analysis of ultrafast single
  skyrmion creation}.
\newblock \emph{\bibinfo{journal}{Phys. Rev. B}} \textbf{\bibinfo{volume}{93}},
  \bibinfo{pages}{174403} (\bibinfo{year}{2016}).
\newblock \urlprefix\url{https://link.aps.org/doi/10.1103/PhysRevB.93.174403}.

\bibitem{PhysRevB.91.224407}
\bibinfo{author}{Lin, S.-Z.}, \bibinfo{author}{Saxena, A.} \&
  \bibinfo{author}{Batista, C.~D.}
\newblock \bibinfo{title}{Skyrmion fractionalization and merons in chiral
  magnets with easy-plane anisotropy}.
\newblock \emph{\bibinfo{journal}{Phys. Rev. B}} \textbf{\bibinfo{volume}{91}},
  \bibinfo{pages}{224407} (\bibinfo{year}{2015}).
\newblock \urlprefix\url{https://link.aps.org/doi/10.1103/PhysRevB.91.224407}.

\bibitem{PhysRevB.83.100408}
\bibinfo{author}{Ezawa, M.}
\newblock \bibinfo{title}{Compact merons and skyrmions in thin chiral magnetic
  films}.
\newblock \emph{\bibinfo{journal}{Phys. Rev. B}} \textbf{\bibinfo{volume}{83}},
  \bibinfo{pages}{100408} (\bibinfo{year}{2011}).
\newblock \urlprefix\url{https://link.aps.org/doi/10.1103/PhysRevB.83.100408}.

\bibitem{DasSarma08}
\bibinfo{author}{Das~Sarma, S.} \& \bibinfo{author}{A., P.}
\newblock \emph{\bibinfo{title}{Perspectives in Quantum Hall Effects: Novel
  Quantum Liquids in Low-Dimensional Semiconductor Structures}}
  (\bibinfo{publisher}{John Wiley and Sons}, \bibinfo{year}{2008}).

\bibitem{PhysRevLett.118.097203}
\bibinfo{author}{Eggebrecht, T.} \emph{et~al.}
\newblock \bibinfo{title}{Light-induced metastable magnetic texture uncovered
  by in situ lorentz microscopy}.
\newblock \emph{\bibinfo{journal}{Phys. Rev. Lett.}}
  \textbf{\bibinfo{volume}{118}}, \bibinfo{pages}{097203}
  (\bibinfo{year}{2017}).
\newblock
  \urlprefix\url{https://link.aps.org/doi/10.1103/PhysRevLett.118.097203}.

\bibitem{Pereiro:2014aa}
\bibinfo{author}{Pereiro, M.} \emph{et~al.}
\newblock \bibinfo{title}{Topological excitations in a kagome magnet}.
\newblock \emph{\bibinfo{journal}{Nature Communications}}
  \textbf{\bibinfo{volume}{5}}, \bibinfo{pages}{4815} (\bibinfo{year}{2014}).
\newblock \urlprefix\url{https://doi.org/10.1038/ncomms5815}.

\bibitem{Yimei18}
\bibinfo{author}{Fu, X.} \emph{et~al.}
\newblock \bibinfo{title}{Optical manipulation of magnetic vortices visualized
  in situ by lorentz electron microscopy}.
\newblock \emph{\bibinfo{journal}{Science Advances}}
  \textbf{\bibinfo{volume}{4}} (\bibinfo{year}{2018}).
\newblock \urlprefix\url{https://advances.sciencemag.org/content/4/7/eaat3077}.
\newblock
  \eprint{https://advances.sciencemag.org/content/4/7/eaat3077.full.pdf}.

\bibitem{Hubert-book}
\bibinfo{author}{Hubert, A.} \& \bibinfo{author}{Schafer, R.}
\newblock \emph{\bibinfo{title}{Magnetic Domains: The Analysis of Magnetic
  Microstructures}} (\bibinfo{publisher}{Springer}, \bibinfo{year}{1998}).

\bibitem{Koog05}
\bibinfo{author}{Lee, K.-S.}, \bibinfo{author}{Choi, S.} \&
  \bibinfo{author}{Kim, S.-K.}
\newblock \bibinfo{title}{Radiation of spin waves from magnetic vortex cores by
  their dynamic motion and annihilation processes}.
\newblock \emph{\bibinfo{journal}{Applied Physics Letters}}
  \textbf{\bibinfo{volume}{87}}, \bibinfo{pages}{192502}
  (\bibinfo{year}{2005}).
\newblock \urlprefix\url{https://doi.org/10.1063/1.2128478}.
\newblock \eprint{https://doi.org/10.1063/1.2128478}.

\bibitem{PhysRevLett.97.177202}
\bibinfo{author}{Hertel, R.} \& \bibinfo{author}{Schneider, C.~M.}
\newblock \bibinfo{title}{Exchange explosions: Magnetization dynamics during
  vortex-antivortex annihilation}.
\newblock \emph{\bibinfo{journal}{Phys. Rev. Lett.}}
  \textbf{\bibinfo{volume}{97}}, \bibinfo{pages}{177202}
  (\bibinfo{year}{2006}).
\newblock
  \urlprefix\url{https://link.aps.org/doi/10.1103/PhysRevLett.97.177202}.

\bibitem{Choe420}
\bibinfo{author}{Choe, S.-B.} \emph{et~al.}
\newblock \bibinfo{title}{Vortex core-driven magnetization dynamics}.
\newblock \emph{\bibinfo{journal}{Science}} \textbf{\bibinfo{volume}{304}},
  \bibinfo{pages}{420--422} (\bibinfo{year}{2004}).
\newblock \urlprefix\url{https://science.sciencemag.org/content/304/5669/420}.
\newblock
  \eprint{https://science.sciencemag.org/content/304/5669/420.full.pdf}.

\bibitem{Templeton97}
\bibinfo{author}{Arrott, A.} \& \bibinfo{author}{Templeton, T.}
\newblock \bibinfo{title}{Micromagnetics and hysteresis as prototypes for
  complex systems}.
\newblock \emph{\bibinfo{journal}{Physica B: Condensed Matter}}
  \textbf{\bibinfo{volume}{233}}, \bibinfo{pages}{259 -- 271}
  (\bibinfo{year}{1997}).
\newblock
  \urlprefix\url{http://www.sciencedirect.com/science/article/pii/S0921452697003098}.
\newblock \bibinfo{note}{Hysteresis Modeling and Micromagnetism}.

\bibitem{Goodenough58}
\bibinfo{author}{Huber, E.~E.}, \bibinfo{author}{Smith, D.~O.} \&
  \bibinfo{author}{Goodenough, J.~B.}
\newblock \bibinfo{title}{Domain‐wall structure in permalloy films}.
\newblock \emph{\bibinfo{journal}{Journal of Applied Physics}}
  \textbf{\bibinfo{volume}{29}}, \bibinfo{pages}{294--295}
  (\bibinfo{year}{1958}).
\newblock \urlprefix\url{https://doi.org/10.1063/1.1723105}.
\newblock \eprint{https://doi.org/10.1063/1.1723105}.

\bibitem{Weiss00}
\bibinfo{author}{Raabe, J.} \emph{et~al.}
\newblock \bibinfo{title}{Magnetization pattern of ferromagnetic nanodisks}.
\newblock \emph{\bibinfo{journal}{Journal of Applied Physics}}
  \textbf{\bibinfo{volume}{88}}, \bibinfo{pages}{4437--4439}
  (\bibinfo{year}{2000}).
\newblock \urlprefix\url{https://aip.scitation.org/doi/abs/10.1063/1.1289216}.
\newblock \eprint{https://aip.scitation.org/doi/pdf/10.1063/1.1289216}.

\bibitem{2LCrCl3}
\bibinfo{author}{Cai, X.} \emph{et~al.}
\newblock \bibinfo{title}{Atomically thin crcl3: An in-plane layered
  antiferromagnetic insulator}.
\newblock \emph{\bibinfo{journal}{Nano Letters}} \textbf{\bibinfo{volume}{19}},
  \bibinfo{pages}{3993--3998} (\bibinfo{year}{2019}).
\newblock \urlprefix\url{https://doi.org/10.1021/acs.nanolett.9b01317}.

\bibitem{Fukamichi02}
\bibinfo{author}{Guslienko, K.~Y.} \emph{et~al.}
\newblock \bibinfo{title}{Eigenfrequencies of vortex state excitations in
  magnetic submicron-size disks}.
\newblock \emph{\bibinfo{journal}{Journal of Applied Physics}}
  \textbf{\bibinfo{volume}{91}}, \bibinfo{pages}{8037--8039}
  (\bibinfo{year}{2002}).
\newblock \urlprefix\url{https://aip.scitation.org/doi/abs/10.1063/1.1450816}.
\newblock \eprint{https://aip.scitation.org/doi/pdf/10.1063/1.1450816}.

\bibitem{Doring68}
\bibinfo{author}{D{\"o}ring, W.}
\newblock \bibinfo{title}{Point singularities in micromagnetism}.
\newblock \emph{\bibinfo{journal}{Journal of Applied Physics}}
  \textbf{\bibinfo{volume}{39}}, \bibinfo{pages}{1006--1007}
  (\bibinfo{year}{1968}).
\newblock \urlprefix\url{https://doi.org/10.1063/1.1656144}.
\newblock \eprint{https://doi.org/10.1063/1.1656144}.

\bibitem{dipole1}
\bibinfo{author}{Vedmedenko, E.~Y.}, \bibinfo{author}{Oepen, H.~P.},
  \bibinfo{author}{Ghazali, A.}, \bibinfo{author}{L\'evy, J.-C.~S.} \&
  \bibinfo{author}{Kirschner, J.}
\newblock \bibinfo{title}{Magnetic microstructure of the spin reorientation
  transition: A computer experiment}.
\newblock \emph{\bibinfo{journal}{Phys. Rev. Lett.}}
  \textbf{\bibinfo{volume}{84}}, \bibinfo{pages}{5884--5887}
  (\bibinfo{year}{2000}).
\newblock \urlprefix\url{https://link.aps.org/doi/10.1103/PhysRevLett.84.5884}.

\bibitem{dipole2}
\bibinfo{author}{Gouva, M.~E.}, \bibinfo{author}{Wysin, G.~M.},
  \bibinfo{author}{Bishop, A.~R.} \& \bibinfo{author}{Mertens, F.~G.}
\newblock \bibinfo{title}{Vortices in the classical two-dimensional anisotropic
  heisenberg model}.
\newblock \emph{\bibinfo{journal}{Phys. Rev. B}} \textbf{\bibinfo{volume}{39}},
  \bibinfo{pages}{11840--11849} (\bibinfo{year}{1989}).
\newblock \urlprefix\url{https://link.aps.org/doi/10.1103/PhysRevB.39.11840}.

\bibitem{Doniach82}
\bibinfo{author}{Garel, T.} \& \bibinfo{author}{Doniach, S.}
\newblock \bibinfo{title}{Phase transitions with spontaneous modulation-the
  dipolar ising ferromagnet}.
\newblock \emph{\bibinfo{journal}{Phys. Rev. B}} \textbf{\bibinfo{volume}{26}},
  \bibinfo{pages}{325--329} (\bibinfo{year}{1982}).
\newblock \urlprefix\url{https://link.aps.org/doi/10.1103/PhysRevB.26.325}.

\bibitem{Giess73}
\bibinfo{author}{Lin, Y.~S.}, \bibinfo{author}{Grundy, P.~J.} \&
  \bibinfo{author}{Giess, E.~A.}
\newblock \bibinfo{title}{Bubble domains in magnetostatically coupled garnet
  films}.
\newblock \emph{\bibinfo{journal}{Applied Physics Letters}}
  \textbf{\bibinfo{volume}{23}}, \bibinfo{pages}{485--487}
  (\bibinfo{year}{1973}).
\newblock \urlprefix\url{https://doi.org/10.1063/1.1654968}.
\newblock \eprint{https://doi.org/10.1063/1.1654968}.

\bibitem{Takao83}
\bibinfo{author}{Takao, S.}
\newblock \bibinfo{title}{A study of magnetization distribution of submicron
  bubbles in sputtered ho-co thin films}.
\newblock \emph{\bibinfo{journal}{Journal of Magnetism and Magnetic Materials}}
  \textbf{\bibinfo{volume}{31-34}}, \bibinfo{pages}{1009 -- 1010}
  (\bibinfo{year}{1983}).
\newblock
  \urlprefix\url{http://www.sciencedirect.com/science/article/pii/0304885383907722}.

\bibitem{Vousden:2016aa}
\bibinfo{author}{Vousden, M.} \emph{et~al.}
\newblock \bibinfo{title}{Skyrmions in thin films with easy-plane
  magnetocrystalline anisotropy}.
\newblock \emph{\bibinfo{journal}{Applied Physics Letters}}
  \textbf{\bibinfo{volume}{108}}, \bibinfo{pages}{132406}
  (\bibinfo{year}{2016}).
\newblock \urlprefix\url{https://doi.org/10.1063/1.4945262}.

\bibitem{Beg:2015aa}
\bibinfo{author}{Beg, M.} \emph{et~al.}
\newblock \bibinfo{title}{Ground state search, hysteretic behaviour and
  reversal mechanism of skyrmionic textures in confined helimagnetic
  nanostructures}.
\newblock \emph{\bibinfo{journal}{Scientific Reports}}
  \textbf{\bibinfo{volume}{5}}, \bibinfo{pages}{17137} (\bibinfo{year}{2015}).
\newblock \urlprefix\url{https://doi.org/10.1038/srep17137}.

\bibitem{Nagaosa13}
\bibinfo{author}{Iwasaki, J.}, \bibinfo{author}{Mochizuki, M.} \&
  \bibinfo{author}{Nagaosa, N.}
\newblock \bibinfo{title}{Current-induced skyrmion dynamics in constricted
  geometries}.
\newblock \emph{\bibinfo{journal}{Nature Nanotechnology}}
  \textbf{\bibinfo{volume}{8}}, \bibinfo{pages}{742--747}
  (\bibinfo{year}{2013}).
\newblock \urlprefix\url{https://doi.org/10.1038/nnano.2013.176}.

\bibitem{Amilcar20}
\bibinfo{author}{Bedoya-Pinto, A.} \emph{et~al.}
\newblock \bibinfo{title}{Intrinsic 2d-xy ferromagnetism in a van der waals
  monolayer}.
\newblock \emph{\bibinfo{journal}{arXiv:2006.07605}} .

\bibitem{Buttner:2017aa}
\bibinfo{author}{B{\"u}ttner, F.} \emph{et~al.}
\newblock \bibinfo{title}{Field-free deterministic ultrafast creation of
  magnetic skyrmions by spin--orbit torques}.
\newblock \emph{\bibinfo{journal}{Nature Nanotechnology}}
  \textbf{\bibinfo{volume}{12}}, \bibinfo{pages}{1040--1044}
  (\bibinfo{year}{2017}).
\newblock \urlprefix\url{https://doi.org/10.1038/nnano.2017.178}.

\bibitem{Schulz:2012aa}
\bibinfo{author}{Schulz, T.} \emph{et~al.}
\newblock \bibinfo{title}{Emergent electrodynamics of skyrmions in a chiral
  magnet}.
\newblock \emph{\bibinfo{journal}{Nature Physics}}
  \textbf{\bibinfo{volume}{8}}, \bibinfo{pages}{301--304}
  (\bibinfo{year}{2012}).
\newblock \urlprefix\url{https://doi.org/10.1038/nphys2231}.

\end{thebibliography}



\pagebreak 


\subsubsection*{Figure captions}

\begin{figure}[htbp]
\centering
\caption{\label{fig1}\textbf{Nucleation of merons and antimerons 
during cooling process.} 
{\bf a-c,} Dynamical spin configurations obtained at different temperatures (T)
showing the evolution of the domain structure and the formation of merons and antimerons 
during field cooling in an external field of 0.0 mT.  
The $S^z$ component is used to follow the evolution of the different spin textures 
across the crystal surface (color map). 
Strong spin fluctuations are observed at temperatures below the critical 
temperature (T$_{\rm c}=$19.07 K, see Supplementary Figure \ref{SI-Tc}) which 
incidentally vanished as the system cools down. 
Localized small areas within 0 K $\leq$T$\leq$ 5 K correspond to spins 
pointed perpendicular to the easy-plane of CrCl$_3$
in different spin polarizations (e.g. $+1$ or $-1$). 
At 0 K, most of the magnitudes of $S_z$ are zero throughout the crystal 
except at well defined small spots with either $S^z=+1$ or $S^z=-1$
in their cores. The formation of 
merons and antimerons occur simultaneously during the time evolution 
without a clear preference over the nucleation site. That is, boundaries, defects or 
edges are not considered. 
{\bf d-f,} Similar as {\bf a-c} but at an external field of 50 mT. 
The applied field polarizes the spin configurations resulting 
in less fluctuations along of $S^z$ even though with alike domain dynamics. 
At T$\leq$ 5 K, the merons and antimerons are still formed but with a 
more preferential spin polarization, e.g. darker spots. If larger magnetic 
fields beyond 50 mT are applied (e.g. 100 mT), a full polarization of the 
topological spin textures is observed with a totality of just one kind of spin polarization. 
For fields above 150 mT, there is no additional nucleation of 
merons and antimerons throughout the crystal as the spin 
textures outside the vortex core follows the field direction. 
See Supplementary Figures \ref{SI-low-field}, \ref{SI-largeB} and 
Supplementary movie S1$-$S3 for details. Scale bar is 50 nm. 
}
\end{figure}

\begin{figure}[htbp]
\centering 
\caption{\label{fig2} {\bf Characterizing the spin features of merons and antimerons.} 
{\bf a,} Snapshot of a spin configuration projected along the $S^z$ component (color map) 
at a selected time with the formation of merons and antimerons 
at zero magnetic field and 0 K. Some of the topological spin textures are 
marked as B1$-$B4 to highlight their features. 
{\bf b,} Profile of $S^z$ along the dashed lines in {\bf a} over the distinct spin quasiparticles  
(B1, B2, B3, B4) showing their widths in \AA's. 
B1 and B2 are close enough to feel the opposite spin polarization from each other. 
The largest peak at 0 \AA ~is centered at the center of the meron or antimeron. B2 and B4 
have both positive spin polarization with similar peak magnitude at 0 \AA ~($S^z=+1$), which is the opposite of B1 and B3. 
{\bf c-f,} Spin textures of the different quasi-particles stabilized in monolayer CrCl$_3$. 
Merons ({\bf c, d}) and antimerons ({\bf e, f}) can be determined by the topological number 
$N$ (Eq.\ref{topological-N}) which involves the vorticity ($\pm 1$) and the core polarization. 
B1 and B4 have the same vorticity of $+1$ even though they are meron ($N=-1/2$) 
and antimeron ($N=+1/2$), respectively. Similar argument applies to B2 and B3, that is, both 
have vorticity of $-1$ but are meron and antimeron, respectively. 
Small arrows (green) indicate the direction and magnitude of the in-plane 
magnetization relative to the core (zero value). 
The underneath color gradient shows the variation of $S^z$ around the spin textures. It reaches 
its maximum magnitude at the core of the merons and antimerons. 
Large arrows (orange and green) give the average behavior observed around the 
vortex core by the in-plane magnetization. 
}
\end{figure}

\begin{figure}[htbp]
\centering
\caption{\label{fig3}\textbf{Meron and antimeron collision.}  
{\bf a-c,} and {\bf d-f,} Snapshots of the vortex and antivortex dynamics 
with antiparallel and parallel core polarization, respectively, before 
({\bf a, d}), during ({\bf b, e}) and after ({\bf c, f}) the collisions at zero field and 0 K. 
The dark and bright backgrounds indicate a more out-of-plane 
magnetization at the core of the vortex and antivortex. 
The big arrows in {\bf a}, {\bf b}, {\bf d} and {\bf e} indicate 
the average behavior of the in-plane spin components in the perimeter. 
In all considered topological spin textures, similar collision scenarios 
are observed which take roughly between 0.57 ns ({\bf a-c}) and 
0.17 ns ({\bf d-f}) to occur. Scale bar is 4 nm. 
{\bf g-j,} Macroscale magnetic domains (blue and red) at different times  
showing the evolution of the merons and antimerons (small circles in green and orange) at 
the boundary between magnetic domains. 
The $S^y$ component (color map) of the magnetization is 
utilized to show the magnetic domain structure whereas $S^z$ for the vortex and antivortex 
textures. The white boundary near where the merons and antimerons are localized 
have $S^y=0$ and $S^x=0$ (see Supplementary Figure \ref{SI-Sx-frustration}). 
$S^z$ reaches its maximum magnitude at the center of the 
spin textures (inset color map in {\bf g}). 
The dynamics of the domains is directly coupled to 
the motion of the merons and antimerons, and vice-versa. 
At sufficient longer times, the entire system results in a 
mono-domain throughout the surface.   
} 
\end{figure}

\begin{figure}[htbp]
\centering
\caption{\label{fig4}\textbf{Dipolar interactions driven the formation of merons and antimerons.} 
{\bf a,} Snapshot of one of the spin dynamics at 0 K and zero magnetic field 
showing the out-of-plane spin component $S^z$ (color map) throughout 
the surface of monolayer CrCl$_3$. {\bf b-d,} Projection of the dipole-dipole interactions 
along of $z$, $x$ and $y$ directions, respectively, on the snapshot in {\bf a}. 
The dipole fields are quantified in mT with positive (red) and negative (blue) 
magnitudes in the color scale. The scale bar of 50 nm is common to all panels.   
}
\end{figure}

\begin{figure}[htbp]
\centering
\caption{\label{fig5}\textbf{Spin fluctuations-driven magnetic domain metastability.} 
{\bf a,} Snapshot of a spin dynamics of monolayer CrCl$_3$ 
obtained through zero-field cooling after 2 ns and reaching 0 K. 
The magnetization perpendicular to the surface (S$^z$) 
is displayed showing the formation of merons and antimerons (small dots). 
Bright (dark) areas correspond to  S$^z=\pm 1$, respectively in the colour scale. 
A path (dashed line) of 200 nm is drawn to show the spatial variation of the 
magnetization at different times ($t \geq 2.20$~ns).  
{\bf b-d,} Variations of the in-plane components of the magnetization (S$^x$, S$^y$) 
and S$^z$, respectively, along of the path shown in {\bf a} within 2.20$-$3.80 ns 
after 0 K is reached. The inset in {\bf d} shows a small area from {\bf a} along the path 
with the formation of the merons and antimerons. The corresponding 
variation of S$^z$ at different times at A, B and C is also showed. 
}
\end{figure}

\pagebreak{}


\subsubsection*{Figures}

\setcounter{figure}{0}

\begin{figure}[htbp]
\centering
\includegraphics[width=1.05\linewidth]{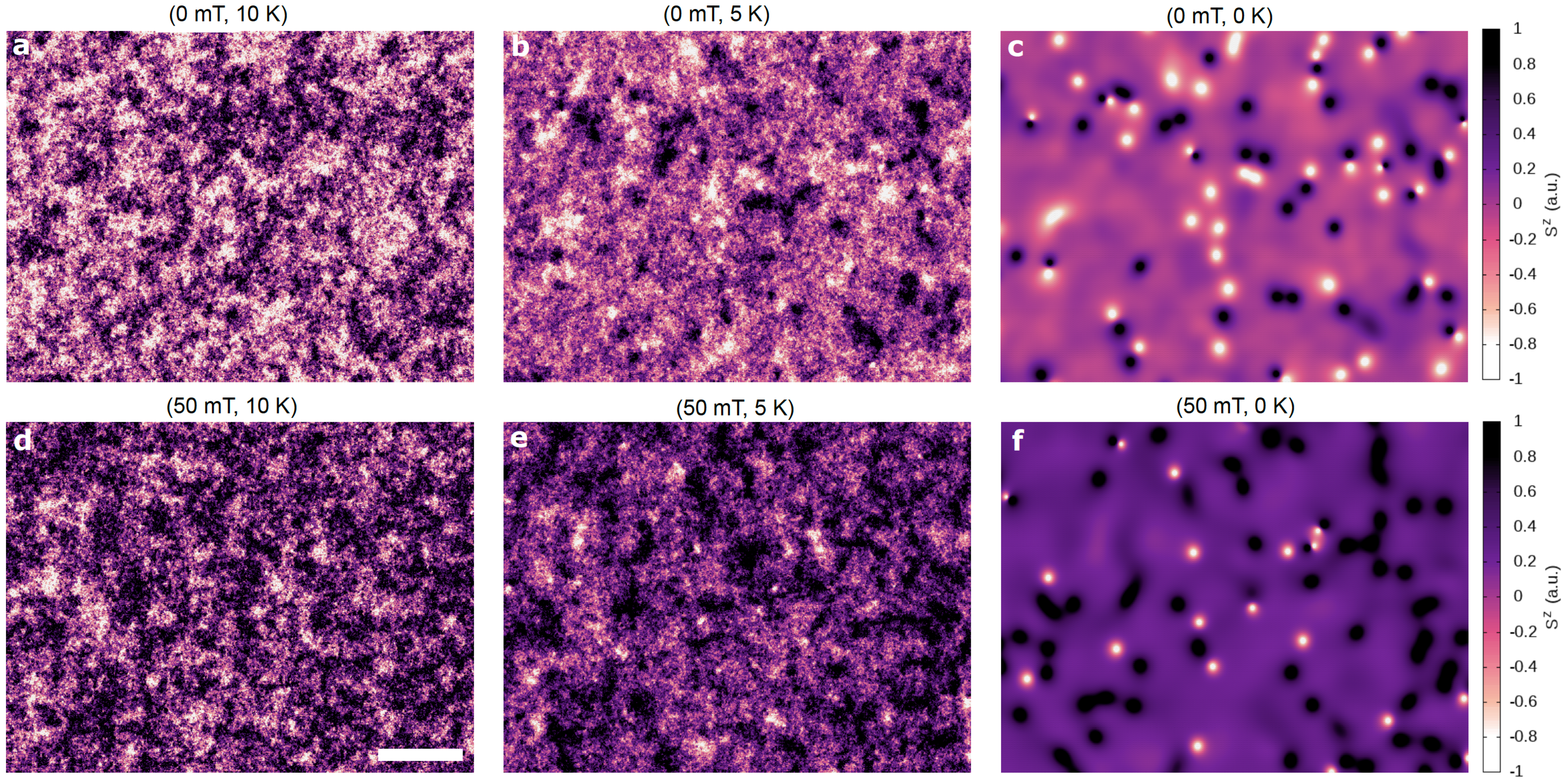}
\caption{}
\end{figure}

\begin{figure}[htbp]
\centering
\includegraphics[width=0.99\linewidth]{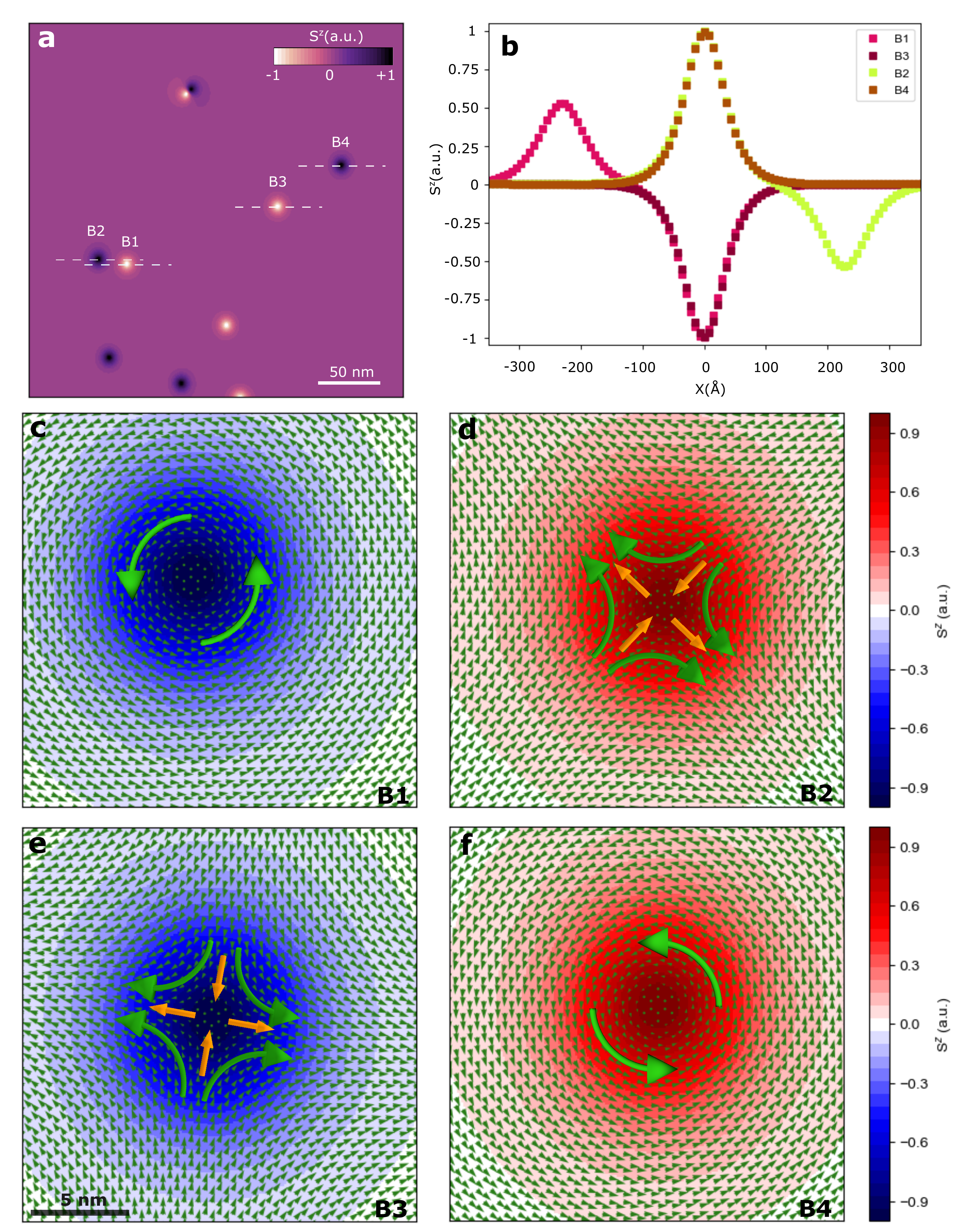}
\caption{}
\end{figure}

\begin{figure}[htbp]
\centering
\includegraphics[width=1.1\linewidth]{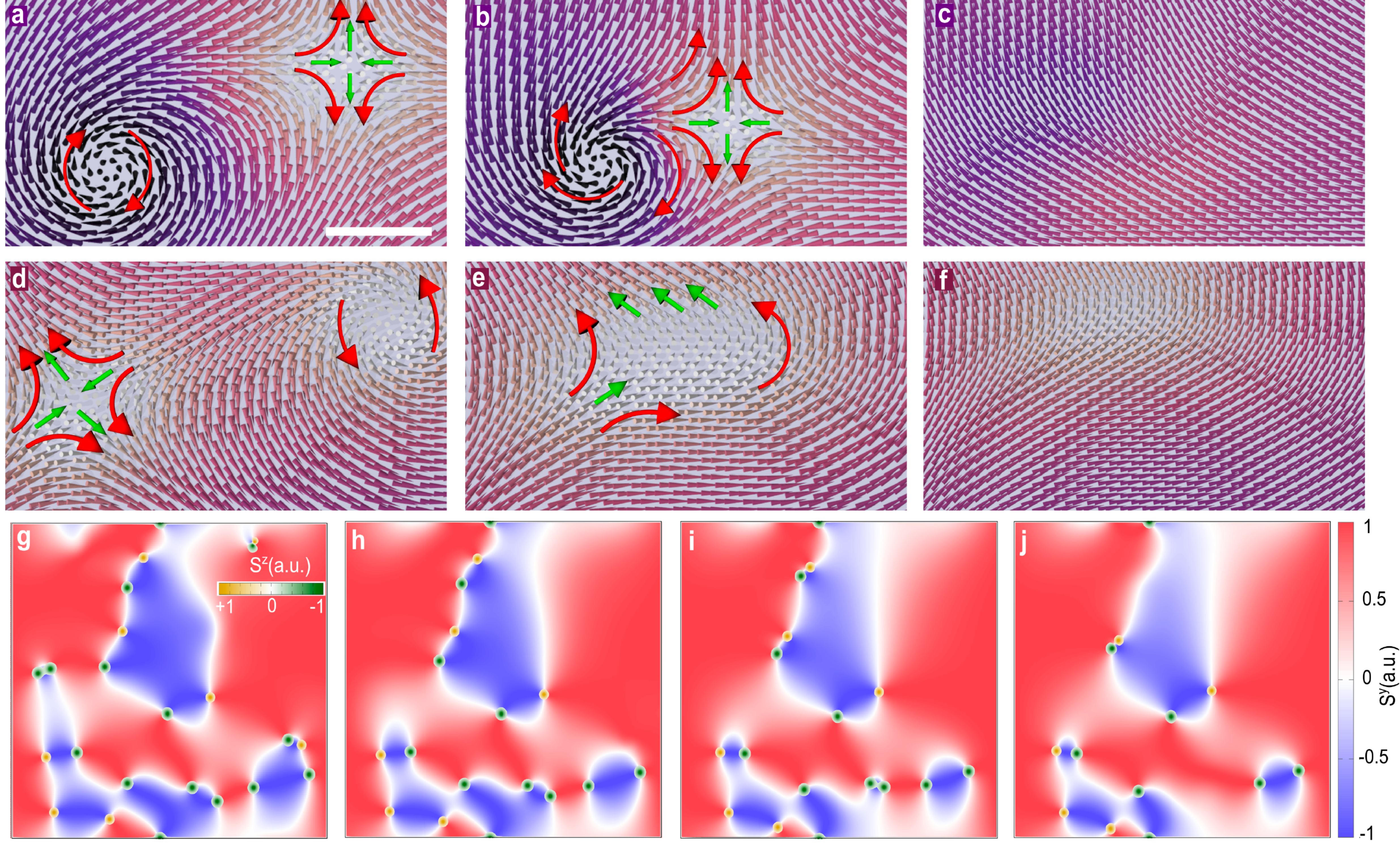}
\caption{}
\end{figure}

\begin{figure}[htbp]
\centering
\includegraphics[width=1.10\linewidth]{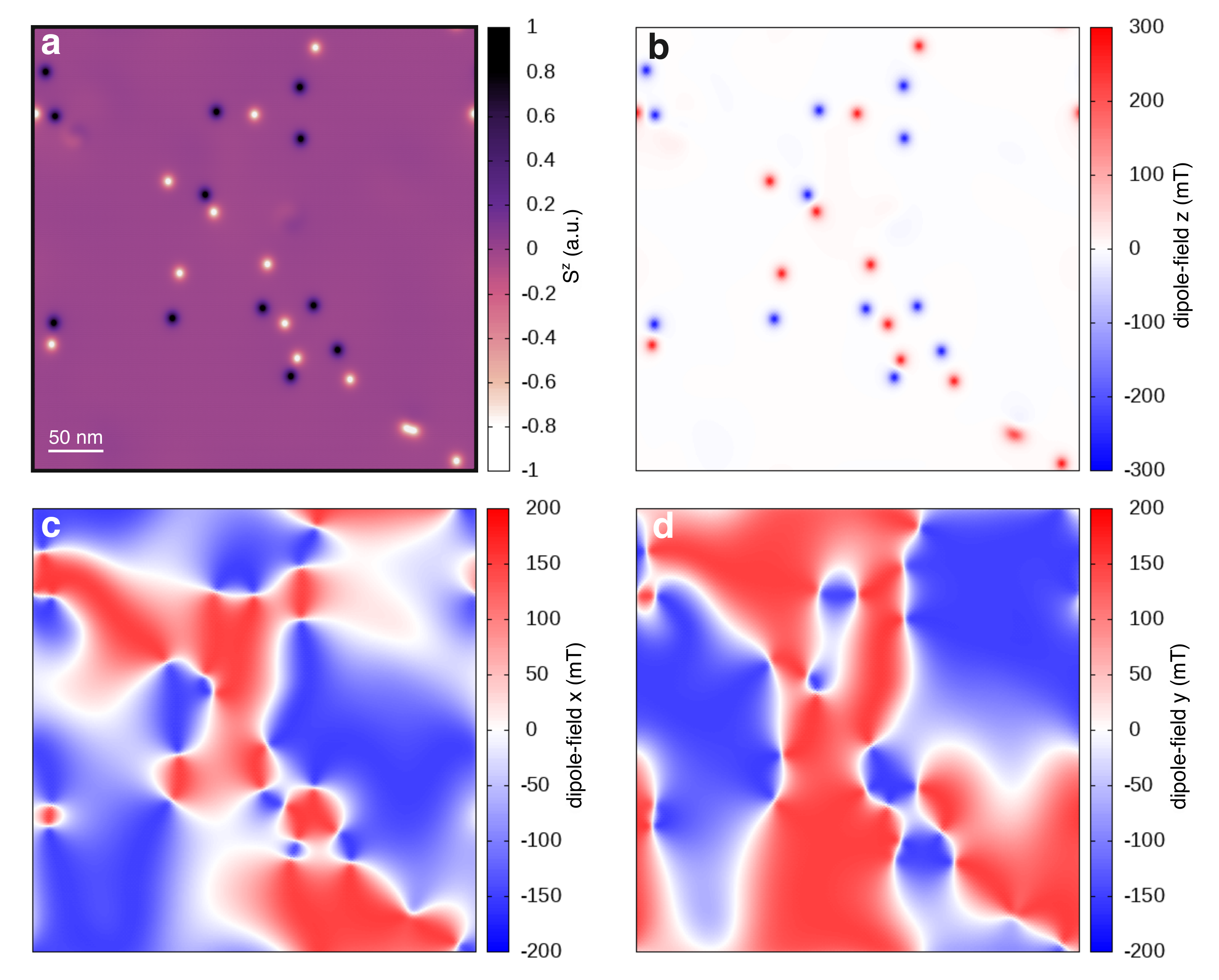}
\caption{}
\end{figure}

\begin{figure}[htbp]
\centering
\includegraphics[width=1.10\linewidth]{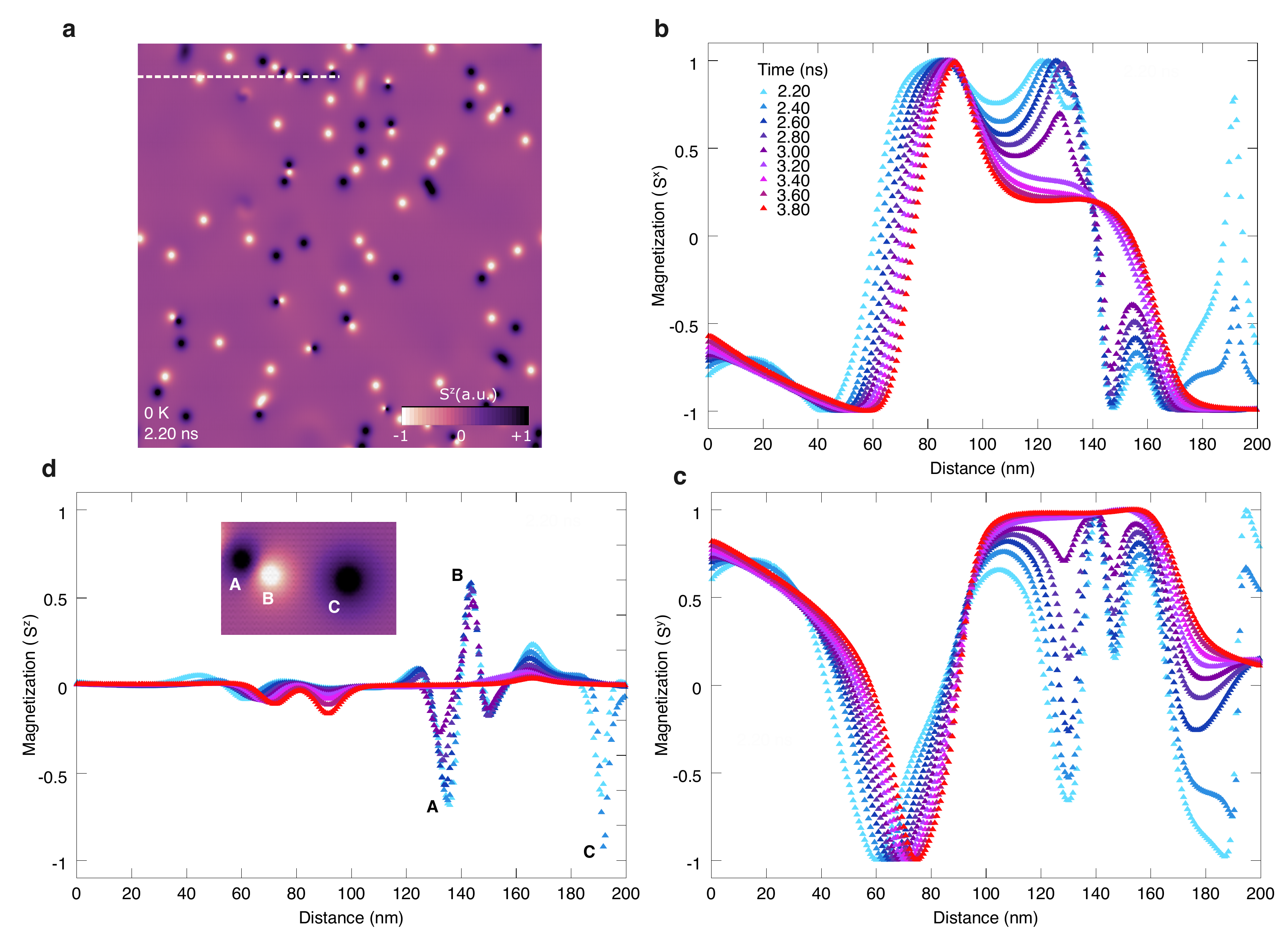}
\caption{}
\end{figure}

\pagebreak{}



\end{document}